\documentclass[manuscript]{emulateapj}
\pdfoutput=1

\usepackage{natbib}
\usepackage{amsmath}
\usepackage{verbatim}
\usepackage{graphicx}
\usepackage{url}
\usepackage[backref,breaklinks,colorlinks,citecolor=blue]{hyperref}
\usepackage[all]{hypcap}

\newcommand{\lya}{Ly$\alpha$}
\newcommand{\oiii}{[O~{\small III}]}

\newcommand{\degs}{$^{\circ}$}

\newcommand{\p}{$p$}
\newcommand{\po}{$p_0$}
\newcommand{\ord}{\texttt{ord}}
\newcommand{\ext}{\texttt{ext}}

\shortauthors{Beck, M. et al.}

\begin{document}

\title{Spectro-polarimetry confirms central powering in a \lya ~nebula at \lowercase{z} = 3.09}

\author{Melanie Beck\altaffilmark{1}, Claudia Scarlata\altaffilmark{1}, Matthew Hayes\altaffilmark{2,3,4}, Mark Dijkstra\altaffilmark{5,6}, and Terry J. Jones\altaffilmark{1}}

\altaffiltext{1}{Minnesota Institute for Astrophysics, School of Physics and Astronomy, University of Minnesota, 116 Church St., Minneapolis, MN 55455, USA, beck@astro.umn.edu}
\altaffiltext{2}{Department of Astronomy, Oskar Klein Centre, Stockholm University, AlbaNova University Centre, SE-106 91 Stockholm, Sweden}
\altaffiltext{3}{Universit\'{e} de Toulouse,  UPS-OMP; IRAP, F-31000 Toulouse, France}
\altaffiltext{4}{CNRS, IRAP, 14 Avenue Edouard Belin, F-31400 Toulouse, France}
\altaffiltext{5}{Institute of Theoretical Astrophysics, University of Oslo, Postboks 1029, 0858 Oslo, Norway}
\altaffiltext{6}{MPI fuer Astrophysik, Karl-Schwarzschild-Str. 1, 85741 Garching, Germany}

\begin{abstract}
We present a follow-up study to the imaging polarimetry performed by \cite{HayesScarlata2011} on LAB1 in the SSA22 protocluster region. Arguably the most well-known Lyman-$\alpha$ ``blob", this radio-quiet emission-line nebula likely hosts a galaxy which is either undergoing significant star formation or hosts an AGN, or both. We obtain deep, spatially resolved spectro-polarimetry of the \lya~emission and detect integrated linear polarization of \textbf{$9$-$13\%\pm2$-$3\%$} at a distance of approximately 15 kpc north and south of the peak of the \lya~surface brightness with polarization vectors lying tangential to the galactic central source. In these same regions, we also detect a wavelength dependence in the polarization which is low at the center of the \lya~line profile and rises substantially in the wings of the profile. These polarization signatures are easily explained by a weak out-flowing shell model. The spectral dependence of the polarization presented here provide a framework for future observations and interpretations of the southern portion of LAB1 in that any model for this system must be able to reproduce this particular spectral dependence.
However, questions still remain for the northern-most spur of LAB1. In this region we detect total linear polarizatin of between 3 and 20\% at the 5\% significance level. Simulations predict that polarization should increase with radius for a symmetric geometry. That the northern spur does not suggests either that this region is not symmetric (which is likely) and exhibits variations in columns density, or that it is kinematically distinct from the rest of LAB1 and powered by another mechanism altogether.  

\end{abstract}


\section{Introduction}\label{sec: intro}

First discovered over a decade ago during the course of deep optical narrowband imaging \citep{Francis1996, Steidel2000}, Lyman-$\alpha$ ``blobs" (LABs) are large, rare, gaseous nebulae in the high-redshift Universe detectable by their extensive \lya~luminosity.  Found predominantly in regions of galaxy overdensities \citep{Palunas2004, Matsuda2004, Prescott2008, Yang2009}, these objects are some of the most promising candidates for the study of ongoing galaxy formation \citep{Mori2006}.  Displaying a range of sizes from tens to hundreds of kiloparsecs and luminosities spanning $\sim10^{43-44}$ erg s$^{-1}$, LABs are reminiscent of high-redshift radio galaxies, yet most are not associated with strong radio sources \citep{Saito2006}. Instead, it seems that LABs are singularly associated with galaxies of one variety or another as even the famed Nilsson's Blob \citep{Nilsson2006}, widely cited as the most overt example of a host-less LAB,  is now believed to be associated with an AGN \citep{Prescott2015}. Other LABs have been associated with an assortment of galaxy populations including Lyman break galaxies (LBGs) \citep{Matsuda2004}, luminous infrared and submillimeter galaxies (SMGs) \citep{Geach2005, Geach2007, Yang2012}, unobscured and obscured quasars (QSOs) \citep{Bunker2003, Weidinger2004, Basu-Zych2004, Smith2009}, as well as starbursting galaxies \citep{Scarlata2009, Colbert2011}.  

Though most LABs seem to have in common a host galaxy or galaxies, the debate over the powering mechanism of the extended \lya~emission remains unresolved in part due to the fact that many of these galaxies seem unable to produce sufficient ionizing flux to light up the surrounding medium \citep{Matsuda2004, SmithJarvis2007}. In addition to photoionization from luminous AGN and/or young stars as a power source \citep{ HaimanRees2001, Jimenez2006, Geach2009, Cantalupo2012}, other possible mechanisms include mechanical energy injected by supernovae winds during powerful starbursts \citep{Taniguchi2000, Scarlata2009}, and radiative cooling \citep{Haiman2000, Fardal2001, DijkstraLoeb2009, Faucher-Giguere2010, Rosdahl2012}. In reality, it is more than likely that LABs are powered by multiple mechanisms simulataneously \citep{Furlanetto2005}. Theoretical studies have shown that polarization of  \lya~photons can be induced by scattering thus providing a potential diagnostic to probe these various powering mechanisms \citep{LeeAhn1998, RybickiLoeb1999, LoebRybicki1999, DijkstraLoeb2008}. 

In partiuclar, we focus our attention on a giant LAB (dubbed LAB1) in the SSA22 protocluster region first discovered by \cite{Steidel2000}.  This nebula is one of the most well-studied with observations ranging from optical to X-ray.  LAB1 is known to be loosely associated with an LBG \citep[C11,][]{Steidel2000, Matsuda2004},  though the peak \lya~surface brightness (SB) is more likely associated with an $850~\mu$m source \citep{Geach2014} with a weak radio counterpart \citep{Chapman2004} as well as associated detections in the near infrared \citep{Geach2007}, all suggestive of a dust-obscured star-forming galaxy leaking \lya~photons which interact with the surrounding medium. Deep integral-field spectroscopy of the \lya~emission has been presented by \citet{Weijmans2010} supporting this conclusion for the brightest regions of LAB1 though they suggest this mechanism is less promising in regions of lower \lya~SB. Additionally, \cite{McLinden2013} perform longslit NIR spectroscopy of portions of LAB1 detecting [OIII] emission in the LBGs C15 and C11 thus determining their systemic velocity.  
\citet[hereafter H11]{HayesScarlata2011} perform narrow-band imaging polarimetry and report low polarization in the central region rising to P=11.9$\pm$2\% within a radius of 7$''$ (45 kpc physical). Coupled with tangential polarization vectors around the central region, they conclude their observations are consistent with powering from an obscured galaxy resulting in scattered \lya~photons by HI.

Due to the observational expense involved, polarization measurements of spatially extended \lya~emission have so far been attempted only three times. In addition to the work of H11, \cite{Prescott2011} present narrow-band imaging polarimetry of a LAB associated with a radio-quiet galaxy at z=2.66 though polarization was not detected.  \cite{Humphrey2013} present the spectro-polarimetry of the gas surrounding the z=2.34 radio galaxy TXS 0211-122 and report low polarization centrally, rising to P=16.4$\pm$4.6\% in some parts of the nebula and conclude that at least a portion of the nebula is powered by the scattering of \lya ~photons produced by the galaxy within.  In this paper we present a follow-up to H11 with the first spectropolarimetric measurement of a radio-quiet LAB.  
In \S\ref{sec: theory} we discuss \lya~radiative transfer, scattering, and polarization basics. In \S\ref{sec: data} we discuss the observations and data analysis. Our methods and initial results are presented in \S\ref{sec: pol}, and in \S\ref{sec: interp} we present a discussion of our results in the context of recent observational work. Finally, in \S\ref{sec: future} we discuss the future of polarization as a diagnostic tool in relation to upcoming space-based polarimeters. 

\section{\lya~Polarization Basics}\label{sec: theory}
\lya~polarization requires photons be scattered imbuing them with a preferential direction or impact angle. Localized (\textit{in situ}) production of \lya~photons, either from stars or gas, is not expected to have significant polarization as these photons will either not scatter sufficiently or have no preferential orientation. In this section we briefly review the necessary physics behind generating a significant \lya~polarization signal. 

The detection of signifiant polarization fraction of \lya~emission depends on two crucial factors: wing vs resonant scattering and Doppler boosting by thermal atoms in the surrounding medium. \lya~is the transition between the first excited and ground states of hydrogen and is a resonant transition, a doublet consisting of two fine-structure lines: $1S_{1/2} - 2P_{1/2}$ and $1S_{1/2}-2P_{3/2}$. The latter transistion can exhibit polarization while the former cannot as scattering through this transition does not retain information on the scattering angle thus producing an isotropized photon. Scattering ``near'' this doublet is called resonant or core scacttering and has been shown to be a superposition of Rayleigh and isotropic scattering producing a minimum level of polarization \citep{Brandt1959, Brasken1998}.  In most astrophysical circumstances \lya~undergoes this type of scattering and the \lya~photons are repeatedly absorbed and re-emitted until they are either destroyed by dust or escape the surrounding neutral medium. However, thermal motions within the gas cause  `partially' coherent scattering where the absorbed and emitted photons are equal only in the rest-frame of the scattering atom. To the outside observer, the \lya~photons  are Doppler boosted with respect to the scattering atom and thus perform a random walk in both frequency and physical space \citep{Neufeld1990, LoebRybicki1999}.  This can cause the \lya~photons to scatter in the wing of the profile where it has been shown that the phase function and degree of polarization are qualitatively consistent with pure Rayleigh scattering \citep{Stenflo1980}. Furthermore, \cite{Stenflo1980} has shown that wing scattering can produce three times more polarization than resonant scattering.  Thus, photons scattering in the wing of the profile are those which are most highly polarized and which see the lowest optical depth in the surrounding medium, enabling them to escape preferentially.

Theoretical predictions have been made by \cite{DijkstraLoeb2008} for the expected amount of polarization in the \lya~line for various astrophysical situations. They explore both an expanding shell and a collapsing cloud. The expanding shell is a simple model of backscattering off a galactic outflow and predicts polarization to increase with radial distance from the central source with total \lya~polarization as high as $40\%$, depending on the assumed column density and velocity of the outflow. Such large values of polarization can be understood due to the kinematics of the gas. Photons scattering off the ``back" of the expanding shell are quickly shifted out of resonance with the gas and into the wing of the line profile thus allowing many to escape after a single wing-scattering.  Similar levels of total polarization (\p$\sim35\%$) are expected in the case of cooling radiation from a collapsing, optically thick gas cloud with the polarization again increasing as a function of radius from the central source due to photons emitted over a spatially extended region within the cloud. . 

In both cases, detecting a high level of polarization through narrow-band imaging polarimetry would be able to rule out \textit{in situ} production of \lya~photons. However, imaging polarimetry alone can not distinguish between outflows or inflows as both predict similar levels of polarization and increasing polarization as a function of radius from the central source. Instead, the frequency dependence of \lya~polarization is required. For the case of an outflowing thin shell, \cite{DijkstraLoeb2008} predict that \lya~polarization will increase redwards of the line center. This is because the redder \lya~photons appear farther from resonance in the frame of the gas and scatter less thus achieving higher levels of polarization. In fact, \cite{DijkstraLoeb2008} state that this frequency dependence can be interpreted as a ``fingerprint" for outflows and predict that \lya~polarization could be as high as $\sim60\%$ in the reddest part of the line profile.  In stark contrast, polarization  increases blueward of line center for a collapsing cloud. Thus the frequency dependence of \lya~polarization can also constrain the kinematic structure of the surrounding gas.

\section{Observations, Reduction, and Calculations}\label{sec: data}
We now turn our attention to the spectro-polarimetry of LAB1.  \cite{HayesScarlata2011} present imaging polarimetry of this nebula in which they find significant polarization increasing as a function of radius from the point of brightest \lya~SB as calculated in Voronoi bins. Furthermore they find polarization vectors which lie tangentially around this central point and conclude that LAB1 is indeed powered by a bright central galaxy obscured from our line of sight. In this portion of the paper we present follow-up spectro-polarimetry in order to confirm and further probe the kinematics of this enigmatic object. In this section we discuss the observations and data reduction methods, as well as the polarization and error calculations performed.

\subsection{Observations}
We choose as our target one of the largest known \lya~blobs located in the SSA22 protocluster region at z=3.09 (see \cite{Steidel2000}).  Dubbed LAB1, this object was observed over the course of five consecutive half nights from 5-9 October 2010, using the FOcal Reducer and low dispersion Spectrograph (FORS2) \citep{Appenzeller1998} instrument mounted on the Antu (UT1) node of the Very Large Telescope (VLT) European Southern Observatory (ESO). The first stage of the dedicated dual-beam polarization optics is the introduction of a strip mask designed to avoid overlapping on the CCD of the two beams of polarized light.  Six MOS slitlets, each 1$''$ wide and 20$''$ long are then positioned over the objects of interest. The light is passed through a super-achromatic half-wave plate (HWP) retarder mosaic (\texttt{RETA2+5}), which rotates the angle of the polarized light. We adopt the standard four angles for unambiguous recovery of the Q and U Stokes parameters: 0\degs, 22.5\degs, 45\degs, and 67.5\degs. The rotated beam is subsequently passed through a Wollaston prism (\texttt{WOLL\_34+13}), which splits the randomly polarized and unpolarized light into two orthogonal, linearly polarized outgoing beams, arbritrarily denoted  the `ordinary' (\ord) and `extraordinary' (\ext) beams. Finally, the beams are passed through a grism dispersion element  (\texttt{GRISM\_1400V+18}) with a central wavelength of 5200 \AA~($\sim$1271 \AA~rest frame), spectral range of 4560-5860~\AA, and dispersion of .63 \AA/pixel. The spectral resolution is approximately 2100 with our 1$''$ slit width. This produces simultaneous \ord~and \ext~spectra which are then projected onto the CCD. Thus each frame consisted of six spectra: three slits contained only sky, two contained stars which were used in the alignment process (see below), and one was positioned over LAB1. We observed LAB1 at a position angle of $-3.09$\degs~in the standard N=0\degs -- E=90\degs~reference frame. 

Throughout the course of each night the sequence of four HWP position angles was observed repeatedly with two full sequences being completed each night. Integration times were 1,800 seconds for each HWP position with the exception of the first night where each exposure had 1,600 and 2,000 seconds of integration time per HWP position for the first and second sequences respectively. Thus at each retarder position angle we obtain a total integration time of 18,000 seconds. 

The entirety of the observing time was classified as clear or photometric with no clouds present on any given night. Since the \ord~and \ext~beams are obtained simultaneously, deviations from photometricity would anyway have no impact on the determination of the Stokes parameters. Observations were taken around the new moon in order to minimize the sky background at bluer wavelengths with moonrise not occurring until after observations were complete each night. Because we observe LAB1 with constant position angle, atmospheric dispersion will vary with airmass over the course of the exposure time. Airmass ranged from 1.1 to 1.53 and was thus well within the FORS instrument's atmospheric dispersion corrector to compensate.  The bulk of the 20 hours of observation time experienced astronomical seeing which varied between 0.5 and 1.0 arcsecond with median seeing at $\sim0.''75$. There were three observations which experienced seeing as high as $1.7''$. These were excluded from the following analysis although their inclusion does not significantly alter our results -- polarization fractions varied only by a few per cent. We thus obtain a total of 37 individual observations for a total of 12,600 seconds of integration.  

\begin{figure}
\begin{center}
\includegraphics[width=3.45in]{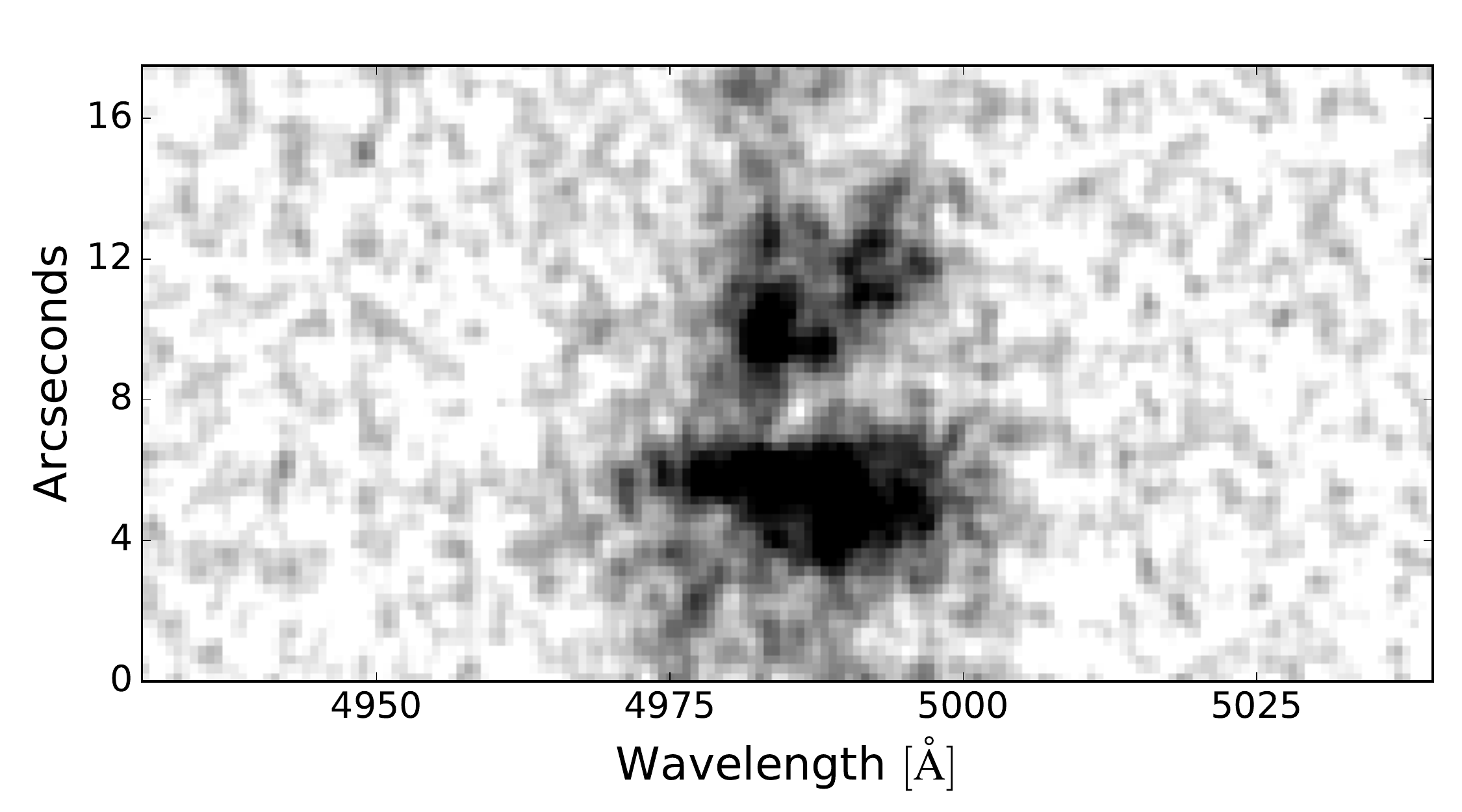}
\caption{Spatial and wavelength distribution of the master total intensity \lya~spectrum. Co-added \lya~spectrum of 77 science spectra smoothed by a Gaussian with FWHM = 0\farcs5. Due to light loss at the edges of the slit, we present here only the central $18''$. }
\label{fig: totint}
\end{center}
\end{figure}


\subsection{Data Reduction}\label{sec: dataredux}

\begin{figure*}
\begin{center}
\includegraphics[scale=1]{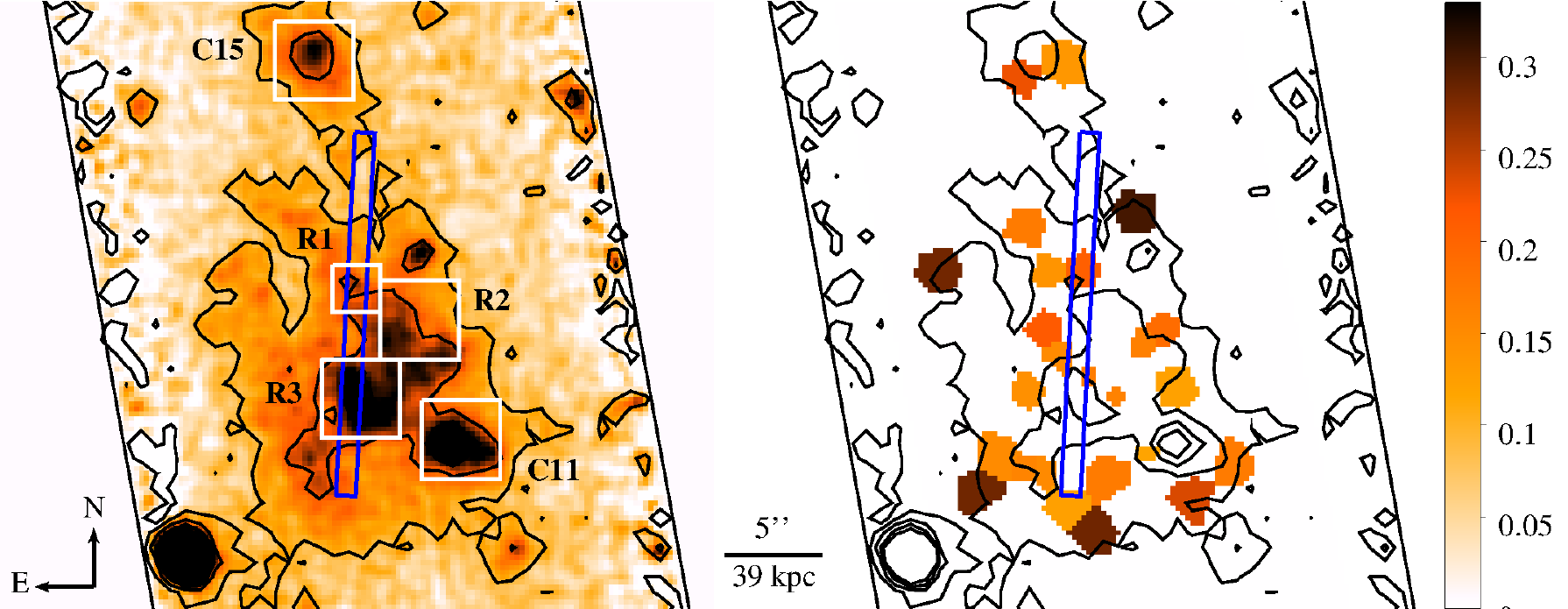}
\caption{Slit position over LAB1 as compared with results from H11. In the left panel we show the combined \lya~intensity frame from H11 adaptively smoothed to show detail including emission from LAB1 and nearby LBGs C11 and C15. Overlaid in white are boxes from \cite{Weijmans2010}, regions in which they performed integral-field spectroscopy on the \lya~emission. As before, the slit shown here spans $\sim$18$''$.  Contours denote arbitrary flux levels. In the right panel we show fractional polarization results from H11 for those bins which were detected at or above 2$\sigma$ (See H11 Fig. 2, panel \textbf{e}). Our slit passes over the region of brightest \lya~emission in the southern portion of the slit as well as a dimmer region in the north.}
\label{fig: slit}
\end{center}
\end{figure*}

The intial steps of data reduction were carried out in the standard manner for spectroscopic observations. Individual frames were biased subtracted. Master flat frames were created from several dome flats using EsoRex\footnote{\url{http://www.eso.org/sci/software/cpl/esorex.html}}, an ESO recipe execution tool, and applied to individual frames to correct for pixel-to-pixel variations. Each frame was then normalized by exposure time. Cosmic rays were throroughly removed using L.A. Cosmic\footnote{\url{http://www.astro.yale.edu/dokkum/lacosmic/}} \citep{vanDokkum2001}. To account for small variations in the spatial direction during observing, frames were aligned using the \texttt{shift\_sub} function in IDL where the pixel shift was calculated as the average of fitted Gaussians of each stellar continua in the slits above and below the science spectra.   At this stage all the individual frames were split into their \ord~and \ext~beams (37 observations $\times$ 2 beams = 74 spectra). Only slits 3 (sky) and 4 (LAB1) were considered for the remainder of the analysis. 

Each beam of the sky and LAB1 spectra was wavelength calibrated individually via a He-Ar arc lamp spectrum using \texttt{NOAO/IRAF}\footnote{IRAF is distributed by the National Optical Astronomy Observatories, which are operated by the Association of Universities for Research in Astronomy, Inc., under cooperative agreement with the National Science Foundation} \texttt{onedspec} and \texttt{twodspec} packages. The \texttt{identify - reidentify - fitcoords - transform} sequence was used on the 2D spectra yielding a fit r.m.s. typically between 0.05 and 0.09 \AA.

Sky subtraction was performed using the sky spectra in slit 3. For each \ord~and \ext~beam, the sky spectrum was median extracted, normalized to the spatial dimension of the 2D LAB1 spectra, and subtracted from the corresponding LAB1 beam.  The residual sky background in the wavelength direction was modeled with a linear fit after masking the \lya~line. This was then subtracted from the spectra. This step was appropriate as there is no evidence of UV continuum according to our preliminary analysis of recently acquired MUSE data which will be published in Hayes et al., in prep. Atmospheric correction was applied using an extinction coefficient as a function of wavelength obtained from \cite{Patat2011} and the airmass at the midpoint of each observation. Spectra were then co-added using a mean combination with simple \texttt{minmax} rejection of the highest and lowest value at each pixel using IRAF's \texttt{imcombine}. As mentioned previously, we exclude 3 observations from further analysis. These include one 45\degs~observation and two 67.5\degs~observations whose delivered seeing was well above $1.0''$. We combine eight groups of 10 spectra to produce science frames for polarimetry (\ord~and \ext~at each of the four angles, see Appendix). We sum each individual frame's \ord~and \ext~beams for a total intensity spectrum for that frame yeilding 40 frames. These are then combined by HWP angle to create four science total intensity frames.  Finally, a master total intensity frame is created by averaging these four frames together, as shown in \autoref{fig: totint}.

\subsection{Polarization and Error Calculations}\label{sec:analysis}
Polarization of \lya~is expected to be linear, thus the decomposition of polarized light falls only into $Q$ and $U$ normalized Stokes parameters. The $V$ parameter represents circular polarization and is not expected for \lya ~radiation. The fourth parameter $I$ is the total intensity which is equal to the sum of the \ord~and \ext~beams.  For each HWP position, $\theta$, the normalized flux difference, $F_{\theta}$, is defined as:

\begin{equation}\label{eqn:normflux}
F_{\theta}=\frac{f_{\theta}^{ord} - f_{\theta}^{ext} }{f_{\theta}^{ord} + f_{\theta}^{ext}}
\end{equation}
where $f^{ord}_{\theta}$ is the flux in the ordinary beam for a given $\theta$ and likewise of $f^{ext}_{\theta}$ for the extraordinary beam.
 
Once the four HWP angles have been obtained, $Q$, $U$, and $I$ relate to the observables by
\begin{equation}
\begin{aligned}
&q = \frac{Q}{I}= \frac{1}{2}F_{0.0} - \frac{1}{2}F_{45.0} \\
&u = \frac{U}{I}= \frac{1}{2}F_{22.5} - \frac{1}{2}F_{67.5}
\end{aligned}\label{eqn:stokes}
\end{equation}
From these, the polarization fraction, \p, and the polarization angle, $\chi$, can be calculated by
\begin{equation}
\begin{aligned}
&p = \sqrt{q^2 + u^2} \\
&\chi = \frac{1}{2} \arctan{\frac{u}{q}}
\end{aligned}\label{eqn:pol-angle}
\end{equation}

However, we actually desire an estimate of the ``true'' polarization,  \po.  When it is assumed that the Stokes parameters $q$ and $u$ are drawn from Gaussian distributions centered around the true values ($q_0$ and $u_0$) each with variance $\sigma$,  it can be shown \cite[e.g.][]{Plaszczynski2014} that the distribution of the polarization follows the Rice distribution:
\begin{equation}
f_p(p) = \frac{p}{\sigma^2}\mathrm{e}^{-\frac{p^2 + p_0^2}{2\sigma^2}}I_0(\frac{pp_0}{\sigma^2})
\end{equation}
where $p_0$ is the true amplitude of the polarization and $I_0$ is the modified Bessel function of the zeroth order.  \autoref{eqn:pol-angle} is considered the naive estimator for this distribution and is known to be strongly biased at low polarization signal-to-noise ratio (SNR$_p$) in part because, in this regime, the Rice distribution can be approximated as a Rayleigh distribution which is highly skewed to larger values of polarization. Additionally, the naive estimator cannot take experimental noise into account. Several attempts have been made to produce an unbiased estimator for \po~\citep[see][for a review]{SimmonsStewart1985} but most have other undesirable qualities such as being unphysical at very low signal-to-noise or containing discontinuities.  \cite{Plaszczynski2014} develop a polarization estimator dubbed the Modifed Asymptotic Estimator (MAS) which is less biased in both low (Rayleigh) and high (Gaussian) SNR$_p$ regimes and is continuous between these regions. Furthermore, this estimator also takes into account measurement error in $q$ and $u$. For these reasons we adopt this estimator for the polarization which goes as:
\begin{equation}\label{eqn: pmas}
\hat{p}_{MAS} = p_i - b_i^2\frac{1-\exp^{-p_i^2/b_i^2}}{2p_i}
\end{equation}
where $p_i$ is given by \autoref{eqn:pol-angle} and $b^2$ is the noise bias of the estimator given by
\begin{equation}\label{eqn: noisebias}
b_i^2 = \frac{q_i^2\sigma_u^2 + u_i^2\sigma_q^2}{q_i^2 + u_i^2}
\end{equation}
where each $i$ is an individual binned measurement of the quantity of interest.

In practice, the quantities $p_i$, $q_i$ and $u_i$ are calculated as given in equations   \ref{eqn:normflux}, \ref{eqn:stokes}, and~\ref{eqn:pol-angle} in each bin (see below for a discussion on our binning strategy).  The $\sigma_q$ and $\sigma_u$ are computed from Monte Carlo simulations whereby each of the eight science frames is allowed to deviate according to a Gaussian spread wherein the deviates are simply computed as the standard deviation of a large portion of the background of each individual science \ord~and \ext~frame. Ten thousand realizations are performed and for each realization $q$ and $u$ are computed. The resulting probability distributions of $q$ and $u$ are Gaussian as expected and from these we obtain $\sigma_q$ and $\sigma_u$ by measuring the spread of the distributions.

We consider the polarization SNR$_p$ by computing 
\begin{equation}\label{eqn: snr}
SNR_p = \frac{p_{MAS}}{\sqrt{\frac{1}{2}(\sigma_q^2 + \sigma_u^2)}},
\end{equation}
where again we allow for measurement error by incorporating both $\sigma_q$ and $\sigma_u$.
Following the prescription of \cite{Plaszczynski2014}, values of SNR$_p > 3.8$ indicate that the Rice distribution is sufficiently Gaussian and thus unbiased. In this regime one may compute a point estimate along with the estimator variance. Values less than this, however, fall in the Rayleigh regime wherein one must instead rely on confidence intervals (CIs).  As we show below, different binning techniques yield different SNR$_p$ and thus in some cases we report point estimates of the fractional polarization while in others we provide the 95\% CIs according to Eqn.~26 in \cite{Plaszczynski2014}.

Finally, we consider the measurement and error estimates for the polarization angle. It has been shown \cite[e.g.][]{WardleKronberg1974, Vinokur1965} that the distribution of $\chi$ is symmetric about the true value of the angle and thus the estimator presented in \autoref{eqn:pol-angle} is already unbiased. At large SNR$_p$ this distribution also tends towards a Gaussian with a standard deviation of  $\sigma_{\chi} \approx \sigma_p/2p$. However, at low SNR$_p$ this approximation underestimates the error. In this work we follow \cite{WardleKronberg1974} and approximate the error by their Eqn.~A6 (see also their Figure 3) which provides the most conservative error estimate for measurements with SNR$_p > 0.5$.

\begin{figure*}
\begin{center}
\includegraphics[scale=.6]{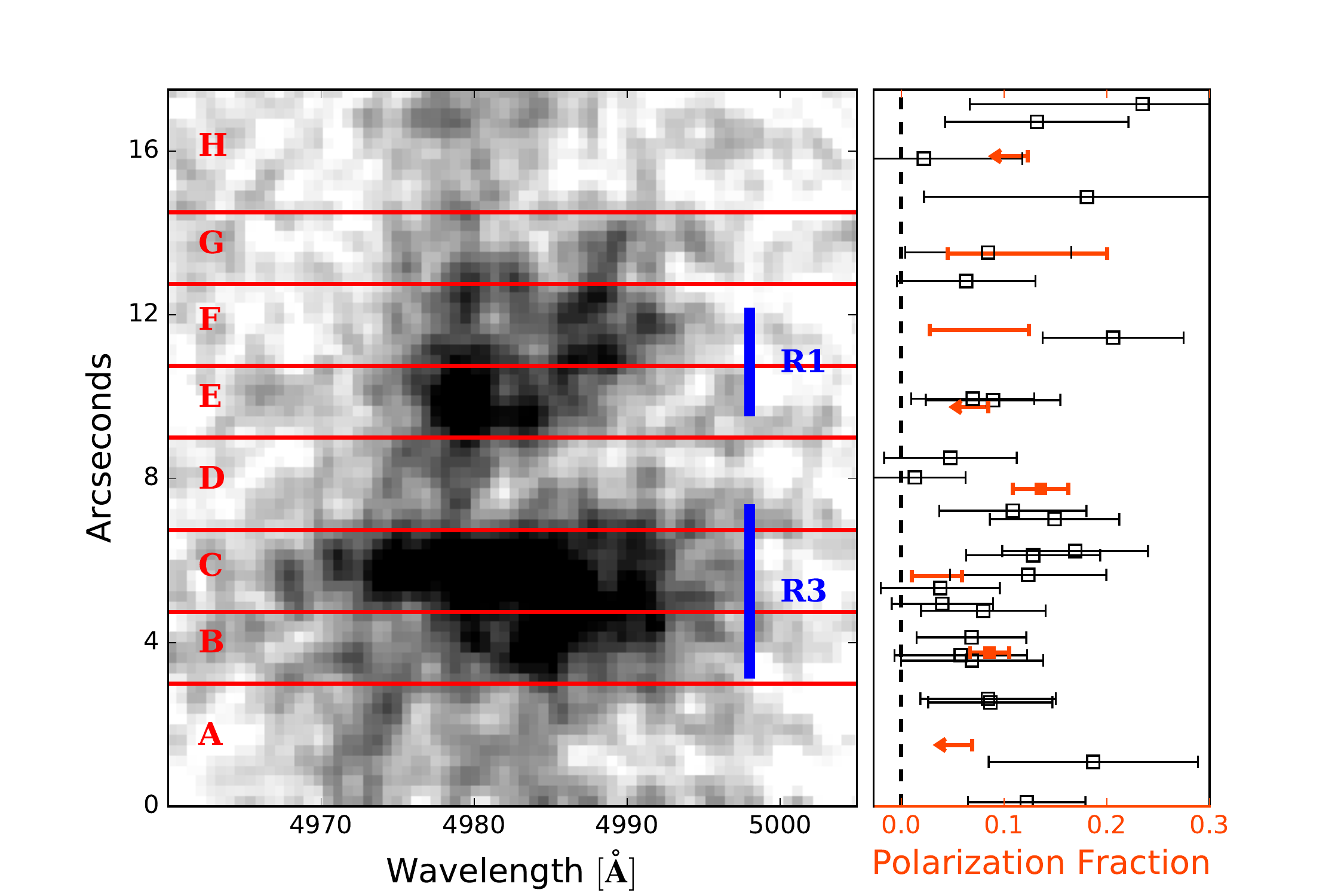}
\caption{Polarization of spectrally integrated \lya. In the left panel we show the spatial binning of the \lya~emission spectrum with bins ranging from 2-3$''$ overlaid in red.  \lya~emission is integrated over 4965-5000~\AA. The blue lines indicate the spatial extent of \cite{Weijmans2010} IFU boxes.  Depicted in the right panel are our polarization measurements in orange.  Polarization point estimates are denoted by orange boxes with $1 \sigma$ error bars;  95\% CIs are shown as closed brackets; and upper limits are denoted with orange arrows where we define our upper limits as the upper 95\% confidence bound for that spatial bin. Black squares show fractional polarization from H11 as measured in Voronoi bins which overlap with our slit. See text for discussion on differences and limitations between datasets.}
\label{fig: totpol}
\end{center}
\end{figure*}

\begin{figure*}
\begin{center}
\includegraphics[scale=.5]{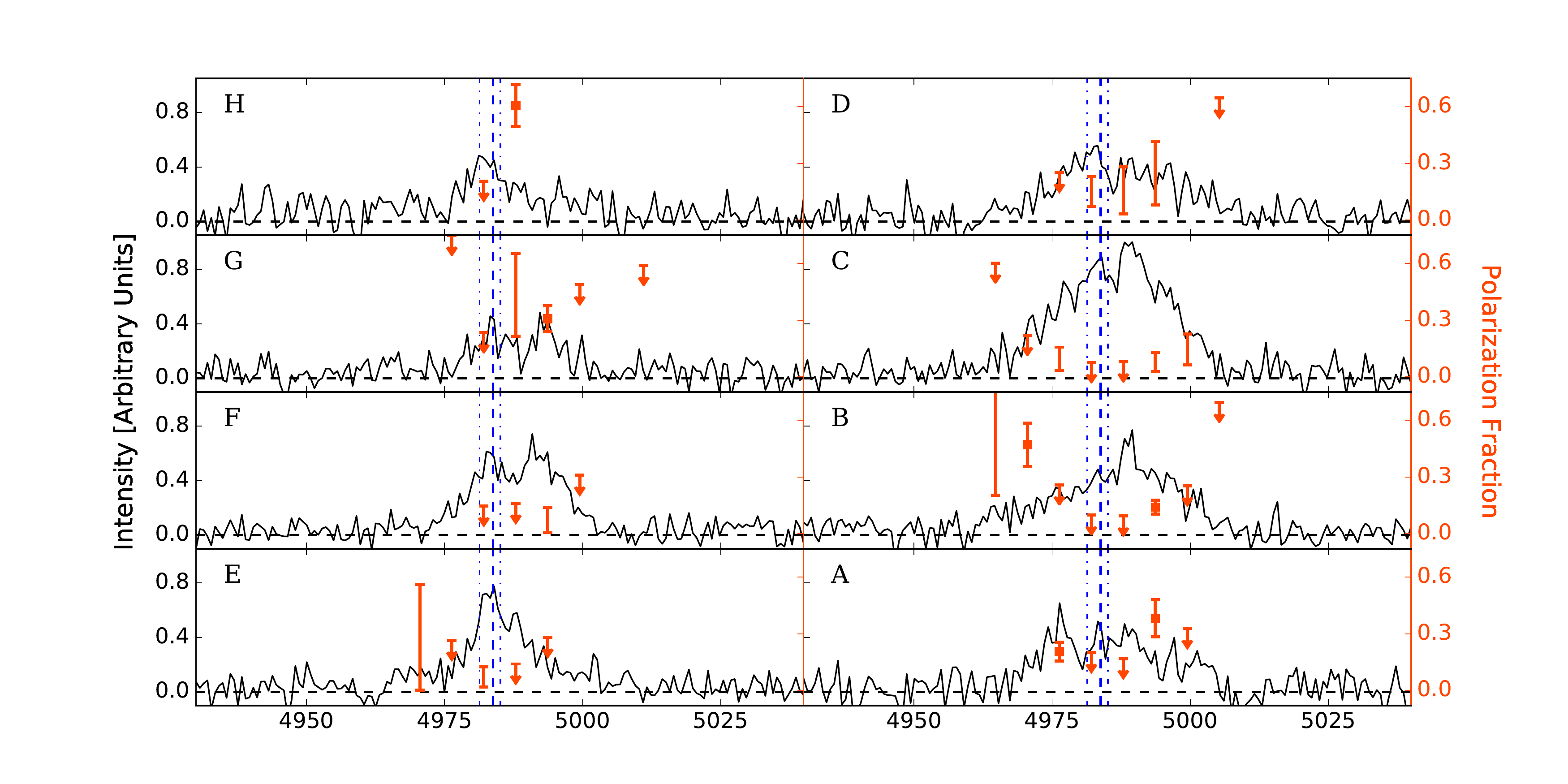}
\caption{Polarization of \lya~as a function of wavelength. Figures \textbf{A} through \textbf{H} contain the extracted 1D \lya~spectrum of the corresponding aperture from \autoref{fig: totpol}. All spectra have been scaled to show their relative intensity. Overplotted in orange we show the polarization fraction as a function of wavelength in bins of 5\AA.  Polarization point estimates are denoted by orange boxes with $1 \sigma$ error bars;  95\% CIs are shown as closed brackets; and upper limits are denoted with orange arrows where we define our upper limits as the upper 95\% confidence bound for that bin. The dashed blue line represents the average systemic velocity as measured from \oiii of four galaxies which are associated with LAB1. The dash-dotted lines are the minimum and maximum of those objects. In bins \textbf{B}-\textbf{E},  we see a trend of low polarization associated with the core of the \lya~emission and higher polarization toward the wings of the line profile. This trend is less apparent in other bins, though the line profiles are not as well defined and many display a double peak. In general, \p~is typically highest at those wavelengths in which the \lya~intensity is relatively low.}
\label{fig: oned}
\end{center}
\end{figure*}

\section{Polarization of \lya~in LAB1}\label{sec: pol}
\subsection{Polarization integrated over the line profile}

Because our data have low signal-to-noise per pixel (SNR$_p \lesssim 1$), binning of the science frames is a necessity. However, any \lya~polarization signal will result from the particular geometry inherent in the HI gas with a unique set of Stokes parameters and polarization angle. If these regions are not azimuthally resolved, one risks overlapping each region's polarization angles thus averaging the polarization signal and potentially washing it out entirely~\citep{DijkstraLoeb2008}. Thus, some binning is necessary but overbinning will make it unmeasurable.

To aid in the determination of appropriate bins we examine the slit position over LAB1 as shown in \autoref{fig: slit}.  LAB1 fills the slit and contains regions of varying \lya~surface brightness (SB) as denoted by the set of arbitrary contours. Also shown in this figure are white boxes corresponding to \lya~emission integral-field spectroscopy as presented by \cite{Weijmans2010}. Our slit overlaps their regions \textbf{R1} and \textbf{R3} and we adopt this nomenclature throughout. \textbf{R3} is situated over the brightest peak of the \lya~SB while \textbf{R1} is associated with a somewhat dimmer region. Between these two features there exists a distinct gap that can also clearly be seen in the \lya~emission shown in \autoref{fig: totint}. We determine to bin these regions separately as they can exhibit different polarization fractions as shown in the right panel of \autoref{fig: slit}. Altogether, we bin the slit into 8 individual spatial regions, each spanning $2''$--$3''$, as this is large enough to achieve adequate SNR$_p$ in some bins yet small enough that we do not wash out any polarization signal. These spatial elements are labelled \textbf{A} through \textbf{H},  with \textbf{A} being the southern-most portion of the slit and \textbf{H} the northern-most. These spatial bins are shown explicitly in \autoref{fig: totpol}.

\begin{table}
	\begin{center}
	\caption{\sf{Polarization signal-to-noise and fractional polarization measurements for spatial bins. For those bins with sufficient SNR$_p$ we report the polarization point estimate and corresonding $1 \sigma_p$ error. Otherwise we report 95\% confidence intervals for bins in which the lower 95\% confidence bound is greater than zero.}}
	\resizebox{.48\textwidth}{!}{
 		\begin{tabular}{l c c c c c}
		\hline
		Bin & SNR$_p$ &  $p_{min}$ & $p_{max}$ & $p$ & $\sigma_p$  \\
		\hline
		\hline
		B &  4.6 &  &  & 8.59 & 1.9\\
		C &  2.6 & 1.00 & 5.92  \\
		D & 4.9  &  &  & 13.6 &  2.7 \\
		F &  2.8  & 2.78 & 12.4 &   \\
		G & 2.8  & 4.51 & 20.0 &  \\
		\hline 
		\end{tabular} }
	\end{center}

	\label{table: bigboxCLs}
\end{table}

With these considerations in mind we first spectrally integrate over the \lya~emission to calculate the total \p~for comparison with H11.  Integration is carried out over the wavelength range 4965--5000~\AA. We note that though the range of the \lya~emission varies within each aperture, varying the integration range only changes the fractional polarization by a few percent difference for all but bin \textbf{H} which has an increase in \p~of $10\%$. However, since we are unable to place reasonable constraints on the polarization fraction in this bin we  consider this to be moot. Only two of these spatial bins have SNR$_p > 3.8$ and for these we report the measured polarization and $1\sigma_p$ error.  The remaining bins have SNR$_p < 3.8$ and for these we report 95\% CIs for those bins in which the 95\% lower confidence bound is greater than zero. Our results are summarized in Table 1 as well as in \autoref{fig: totpol} along with the spatial binning pattern and the polarization measurements from H11 for those bins which our slit overlapped. The fractional polarization values roughly agree when one takes into account the distinct methods used between our two analyses. H11 utilize a Voronoi binning technique whereby the size of each bin is determined by the achieved SNR within that bin. This allows them to have bins of various sizes. Each of their bins only partially overlaps our slit and we include in \autoref{fig: totpol} all such bins. The largest disparity between datasets occurs in Bin F. In this region, H11 measure the fractional polarization to be $p\sim20\%$ whereas we find, at most, $\sim12\%$. The reason for this discrepancy is not fully understood. In \autoref{table: bigboxCLs} we report the 95\% confidence intervals for those spatial bins whose lower 95\% confidence bound is greater than zero. We see polarization in spatial bin \textbf{C} which is on the order of a few per-cent though not quite consistent with zero. The fractional polarization then increases to 13.6\% north and 8.6\% south of this region as seen in spatial bins \textbf{B} and \textbf{D}. The distance between bin \textbf{C} and these bins is roughly 15 kpc in either direction. 

\subsection{Polarization across the line profile}
\begin{figure}
\begin{center}
\includegraphics[width=3.65in]{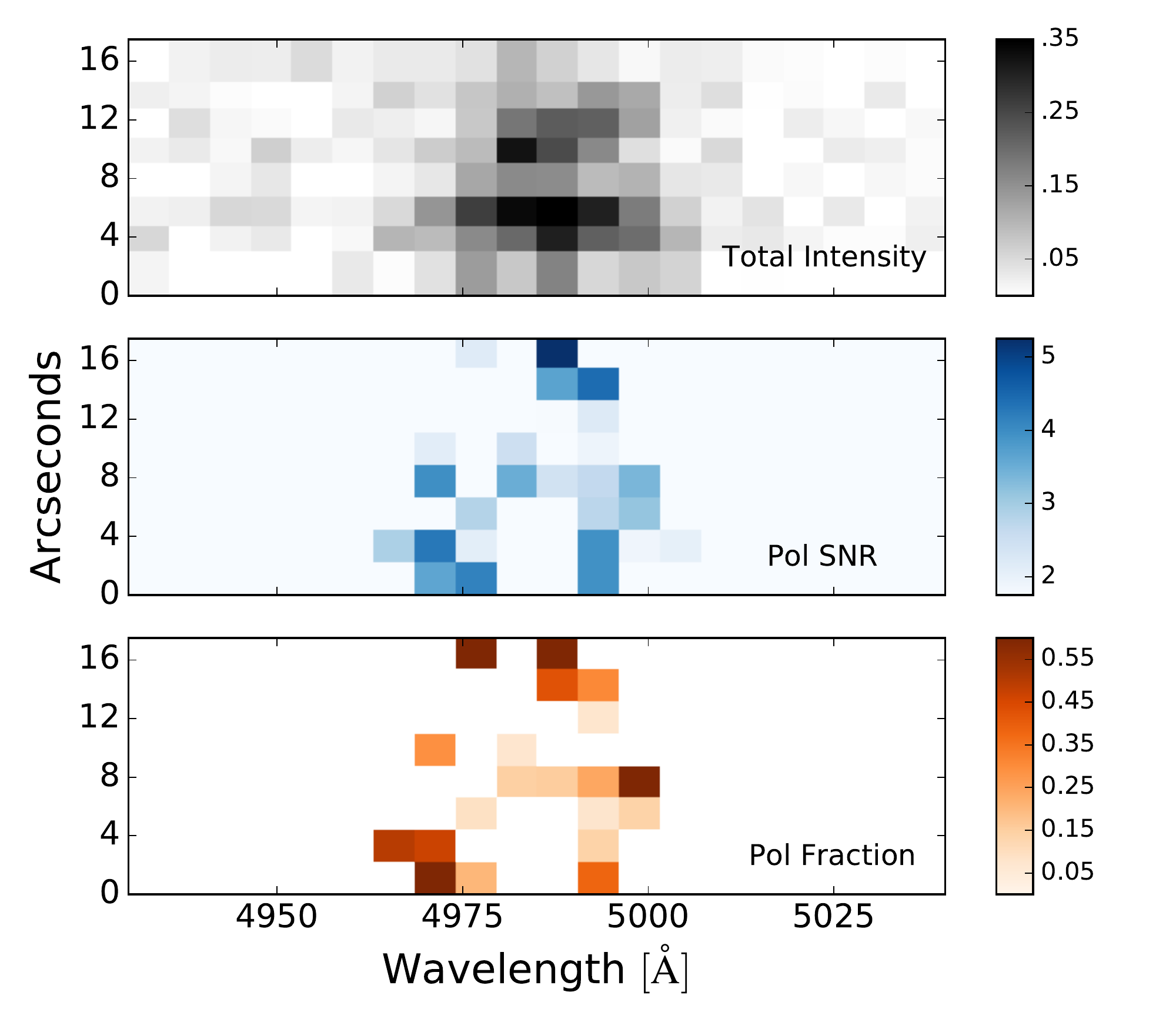}
\caption{\textit{Top.}~Total intensity of the 2D \lya~spectrum in bins of 5\AA~by 2-3$''$. \textit{Middle.}~SNR$_p$ map which demonstrates the relative quality of our polarization measurements. SNR$p > 2$ is generally enough for us to discriminate \p~statistically signicant from zero with 95\% confidence and we show CIs for such bins in \autoref{fig: oned}.  SNR$_p > 3.8$ indicates the Rice distribution is sufficiently Gaussian and we measure polarization with standard errors, also shown in \autoref{fig: oned}. The bottom panel depicts the map of \p~determined to be statistically significant from zero with 95\% confidence (lower 95\% confidence bounds greater than zero). For those bins with sufficient SNR$_p$, the bin color reflects the point estimate of \p, otherwise it  represents the middle value of the CI associated with that bin.}
\label{fig: pol2d}
\end{center}
\end{figure}

We next explore \p~as a function of wavelength. Using the same spatial elements, we further bin each into 5~\AA~increments in the wavlength direction (see the Appendix for comparison of $q$, $u$ and \p). In \autoref{fig: oned} we present the extracted 1D spectra for each spatial element along with the wavelength response of \p~in orange.  For those bins with sufficient SNR$_p$ (as shown in the middle panel of \autoref{fig: pol2d}), we show the polarization point estimate along with $1 \sigma_p$ errors. For those bins which have lower 95\% confidence bounds greater than zero, we show the CI as orange closed brackets. Otherwise, we show upper limits defined as the upper 95\% confidence bound for that bin. In general we see a trend of high (low) polarization corresponding to lower (higher) relative \lya~intensity. In particular, bin \textbf{B}  displays \p~which is consistent with zero in near 4985\AA~but which rises substantially in the wings of the profile, reaching up to $45\%$ bluewards and with upper limits as high as $65\%$ redwards.  In spatial bins \textbf{C} and \textbf{D} we see the suggestion of similar behavior with lower polarization in the core of the line and potentially higher polarization in the wings of the profile though the data do not allow us to furher constrain the trend. 

It is important to recall that these measurements inform us as to the fraction of the total intensity which is polarized. In the case of box \textbf{B}, for example, the wavelength bin at 4970\AA~is 45\% polarized. The intensity in this portion of the line is quite low, however, relative to the peak at 4990\AA. It is instructive to compare this to the integrated polarization in \autoref{fig: totpol}. In that figure, box \textbf{B} has fractional polarization of $\sim9\%$. Thus we see that spectropolarimetry gives us more information than what can be gained solely through imaging or integrated polarimetry. Highly polarized individual wavelength bins are `washed out' by the relative strength of the core of the profile which is typically not strongly polarized. The overall fraction of polarized photons decreases when integrating over the entire line and it is impossible to reconstruct from imaging alone which wavelength regimes are the most highly polarized.

In \autoref{fig: pol2d} we present 2D maps of the spectrally binned intensity, SNR$_p$, and polarization of LAB1. In the middle panel we show the SNR$_p$ where we stress that the ``noise" in this equation is not the equivalent of a polarization error. This figure instead serves to give the reader a feeling for the relative quality of our measurements. Comparing the middle and top panels we see that many areas of high intensity have very low polarization signal-to-noise indicating that these regions most likely have very low or no fractional polarization for us to detect with the current data. However, portions of the spectrum exhibiting relatively less intensity have much more significant SNR$_p$. The fraction of polarized light in these bins is relatively more substantial. We also point out that in this SNR$_p$ map we see a range of values from $\sim2$ to $\sim5$ which indicates that for some bins we report point estimates of the fractional polarization while for others we instead provide 95\% CIs. In the bottom panel, we show \p~in individual bins in which we either have sufficiently high SNR$_p$ or in which we calculate a lower 95\% confidence bound greater than zero.  In general we see higher values of \p~in the reddest part of the line though we also note some substantial polarization in the blue wing as well. In most cases we see that the central emission region is characterized by low SNR$_p$ and low fractional polarization.

\begin{figure}[t]
\begin{center}
\includegraphics[width=3.25in]{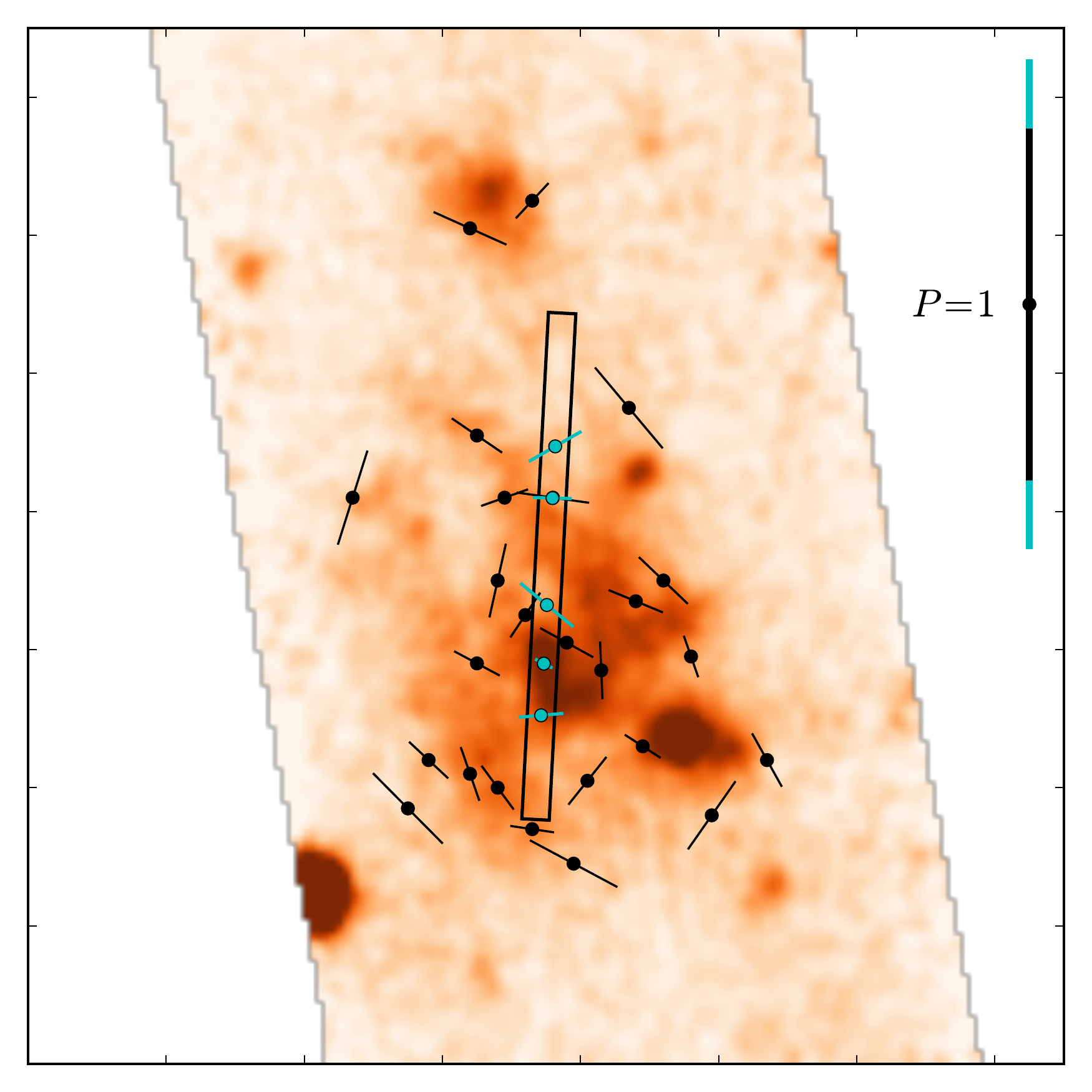}
\caption{Direction of \lya~polarization vectors. Smoothed total \lya~intensity from H11 overlaid with polarization vectors from H11 (black) and from this analysis (cyan) for those spatial bins from \autoref{fig: totpol}~which have SNR$_p > 2$. Because we measure small polarization amplitude in our bins we use a larger scaling for our vectors in order for the reader to more easily compare the angles we measure with with those presented in H11. Both sets of angles generally lie tangentially about the region of highest \lya~SB.}
\label{fig: angles}
\end{center}
\end{figure}

Finally, in \autoref{fig: angles} we present the direction of the polarization vectors, $\chi$, for those spatial elements (\textbf{B-D, F, G}) which have SNR$_p > 2$.   Errors on the polarization angles range from $\sim$6--15\degs.  In this figure we also plot polarization vectors from H11 for comparison, including only those which they measured at or above 2$\sigma$. We see that both sets are generally consistent. Like H11, our angles lie tangentially around the peak of \lya~SB. 

\section{Discussion and Conclusions}\label{sec: interp}

We have presented deep spectro-polarimetry of  the LAB1 \lya~emission nebula. The data allow us to probe the kinematics and distribution of the neutral gas and reinforce the idea that LAB1 is likely composed of several smaller, more complex regions instead of one large, kinematic structure. In particular, our observations suggest at least a weak outflow in the southern portion of the LAB as we discuss below.

Simulations predict that polarization due to scattering should exhibit a radial dependence on the sky. The region of highest \lya~SB would not be strongly polarized but the polarization would rise with increasing radius \citep{DijkstraLoeb2008}.  The observed polarization of \lya~in the southern portion of the slit is consistent with \lya~photons produced by a luminous galaxy (or galaxies) and scattered at large radii by the surrounding neutral hydrogen. As in H11, we see this signature here, most notably in spatial elements \textbf{B-D} where the peak of the \lya~SB has little observable polarization as shown in spatial element \textbf{C}. North of this location  (\textbf{D}),  we find \p~= $13.6\pm2.7\%$ which is also consistent with the Voronoi bins from H11 lying on either side of our slit at approximately the same radial distance (see the right side of \autoref{fig: slit}).  Similarly to the south (\textbf{B}), we find total \p~= $8.59\pm1.9\%$.  Thus we have increasing polarization away from the peak \lya~emission with a radius of $\sim15$ kpc. However, this general trend cannot determine between inflowing or outflowing gas as both are predicted to have this observational signature.

\cite{DijkstraLoeb2008} also predict that, for an envelope of expanding gas, those photons in the bulk of the line profile should exhibit increasing polarization redward of the line core. The strength of this increase depends strongly on the distance from the center of \lya~SB as well as the column density and outflow velocity. In other words, for a given column density and outflow speed, polarization will increase weakly (up to $\sim10\%$) in the reddest part of the wing at the peak of \lya~SB, but will increase substantially (up to $\sim70\%$) at larger radii from this region.  We caution that direct comparison of our data with these simulations comes with a caveat since  we are not integrating over the entire shell with our long-slit observations. Nevertheless, the spectrally resolved polarization in bins \textbf{C} and \textbf{B} at least suggest this behavior.

To see this trend, we first estimate the systemic velocity of the region as the average from four galaxies measured in \oiii~and known to be components of LAB1 \citep{Kubo2015, McLinden2013} (see \autoref{fig: oned}).  We note that these measurements are all within $\sim230$ km/s. Given the width of the \lya~profile, this uncertainty has minimal impact. In this context we can see that the polarization in \textbf{C} is well constrained in the red wing to be at most $\sim20\%$ polarized. Though we are unable to tightly constrain the red wing in bin \textbf{B}, large polarization values are suggested by the detection at 4995\AA, and are not ruled out at longer wavelengths.  This particular polarization pattern can be explained easily with a weak outflowing shell model whereby \lya~photons emitted from an embedded source interact with the expanding shell. Those photons which interact with the receding portion of the shell (from our point of view) are Doppler shifted into the red wing of the profile. Being in the wing of the line profile, these photons see a lower optical depth and thus preferentially escape the medium having only scattered a few times which preserves their polarization. Photons which remain in the core of the profile in the frame of the gas scatter many more times, effectively erasing any polarization signal.

The polarization data presented here provide a framework for future observations of the southern portion of LAB1. The imaging polarimetry of H11 coupled with a recent strong submillimeter source within 1.5$''$ of the peak \lya~intensity \citep{Geach2014} provide the `smoking gun' that there is indeed at least one powerful source embedded in this region, the photons from which are likely scattering at large radii. If there are indeed multiple  sources within about 30 kpc of this region, observations must also provide a mechanism by which these sources can reproduce the spectral polarization signature presented here, namely, a configuration which is consistant with low polarization in the core of the \lya~profile and high polarization in the wings.

However, the picture remains obscure for \textbf{R1} corresponding to our spatial elements \textbf{E-F}. Were \textbf{R1} to be part of the same smooth, kinematic structure as \textbf{R3} we would expect its total \lya~polarization to increase relative to \textbf{R3} due to its increased radius from the galactic center. Instead, we see the total polarization drop and flatten across \textbf{R1}. It's possible that the gas in this region is clumpier or denser than the southern portion of the blob. An increase in the column density of the gas will decrease the observed polarization fraction as additional scatterings tend to isotropize the photons. Another possibility is that this region is powered by flourescence from ionizing radation eminating from the central source. This would naturally explain the lower polarization as this type of \textit{in situ} production of photons is not expected to be highly polarized.   \cite{Weijmans2010} present compelling evidence that suggests this region is kinematically distinct from the rest of LAB1 and thus a third possibility is that this region is instead powered by radiative cooling. Most likely is the possibility that this region is dominated by an embedded souce of its own. Though interesting to speculate, the wavelength dependence of the polarization in these spatial bins is not sufficient for us to further probe the kinematics and polarization properties.


\section{The Future of \lya~Polarization}\label{sec: future}
With another successful detection of the spectral dependence of \lya~polarization the question arises: What does the future hold for \lya~polarization? Additionally, should emphasis be placed on imaging or spectral polarimetry? The integration times for either mode are similar in magnitude and require a substantial commitment so the choice between methods is not a trivial one.  

We have explicitly demonstrated that much information can be gleaned from the spectral dependence of the polarization signal. In particular, features emerge which narrowband imaging polarimetry simply cannot detect. Not only are we able to detect the wavelength dependence for many of our spatial bins but we also find high polarization upwards of 60\% in portions of the \lya~profiles -- information which is completely lost in imaging polarimetry. While imaging polarimetry can confirm the presence of scattering and probe the overall geometry of the scattering medium, it cannot probe the kinematics of the system to determine potential outflows or inflows. Furthermore, the spectrally integrated polarization can still provide spatial clues as to any existing radial dependence with advantageous slit placement. Because the geometry of the blob is important to the overall detection of a polarization signal, spectropolarimetry should not be conducted blindly but instead by guided by spatial information obtained from the already existing narrowband surveys of LABs as well as IFUs. 

Though it remains to be seen, suggestions have arisen that the next generation of $\sim$30 m telescopes could extend \lya~polarization studies. This is supported in that all projected Extremely Large Telescopes (ELT) have proposed polarimetry as a necessary part of their instrument suite.  On the E-ELT, the \textit{Exo-Planet Imaging Camera and Spectrograph} \citep[\textit{EPICS},][]{Kasper2008} includes the \textit{EPOL} polarimeter \citep{Keller2010}.  The Thirty Meter Telescope has discussed plans to include the \textit{Second-Earth Imager for TMT} \citep{SEIT}. And the Giant Magellan Telescope has discussed spectro-polarimetric capabilities as necessary to meeting their science goals \citep{GMTscience}.  Off the ground, several probe-scale NASA space missions have been proposed to study exoplanetary systems such as the AFTA and ``EXO" missions \citep{Stapelfeldt2014, Seager2014}, with considerable emphasis given to polarimeters to enrich the science output. While many of these projects focus on imaging polarimetry, the UVMag consortium has proposed the Arago space mission which would be devoted to unprecedented spectropolarimetry from the FUV through the NIR \citep{Pertenais2014}.  This field is currently driven almost exclusively by exo-planetary science for studying the scattering off planetary atmospheres and circumstellar disks but it is the development of such instrumentation which is of greatest importance. At this stage, we cannot say whether we will be able to point one of these instruments directly towards a \lya~blob without slight modification of the initial design or incorporation of additional settings but it is optically plausible \citep{Hayes2011}.

In the meantime, much can still be accomplished from the ground with 8 m class telescopes. To date, only three \lya~emitting sources have been studied in depth: LAB1 \citep[][and this work]{HayesScarlata2011}, LABd05 \citep{Prescott2011} and radio galaxy TXS 0211--122, known to be associated with a 100 kpc scale \lya~nebula \citep{Humphrey2013}. Radio-quiet LABs were first targeted due to their apparently controversial nature unlike high redshift radio galaxies (HzRGs) which did not pose an energy problem.  With the discovery that a HzRG is at least partially polarized due to \lya~scattering, it behooves us to test further what relationship, if any, exists between radio-loud and -quiet nebulae. Compact \lya~sources also remain unexplored with the polarimeter. Though targeting resolved objects ensures that the Stokes parameters do not all cancel, symmetry is likely broken in most systems. Thus we may expect a measureable signal from LAEs \citep{LeeAhn1998}.
 
Additionally, the interpretation of \lya~polarization is still a challenging prospect of its own. Most state-of-the-art simulations assume density and kinematic structures that are still unrealistic in that variations proceed smoothly. What is urgently needed is the implementation of \lya~polarization in all \lya~radiative transport codes to generate predictions for various applications including clumpy and filimentary media as well as non-spherically symmetric geometries. While some work has already been done in this area \citep{DijkstraKramer2012}, there is still much to be explored in terms of predicting observable polarized \lya~line profiles. With the current limitation on instumentation coupled with exacting observations, we need dedicated theoretical and observational developments that proceed in tandem. 
\newline

We thank the anonymous referee for the useful comments that significantly improved the analysis and presentation of our results. Additionally, MB and CS are grateful to J\'{e}r\'{e}my Blaizot and Maxime Trebitsch for proofing the manuscript and providing feedback which helped to clarify the text. 

\bibliographystyle{apj} 
\bibliography{apj-jour,lyabib}


\appendix\label{sec: app}

\begin{figure}[h]
\begin{center}
\includegraphics[width=7in]{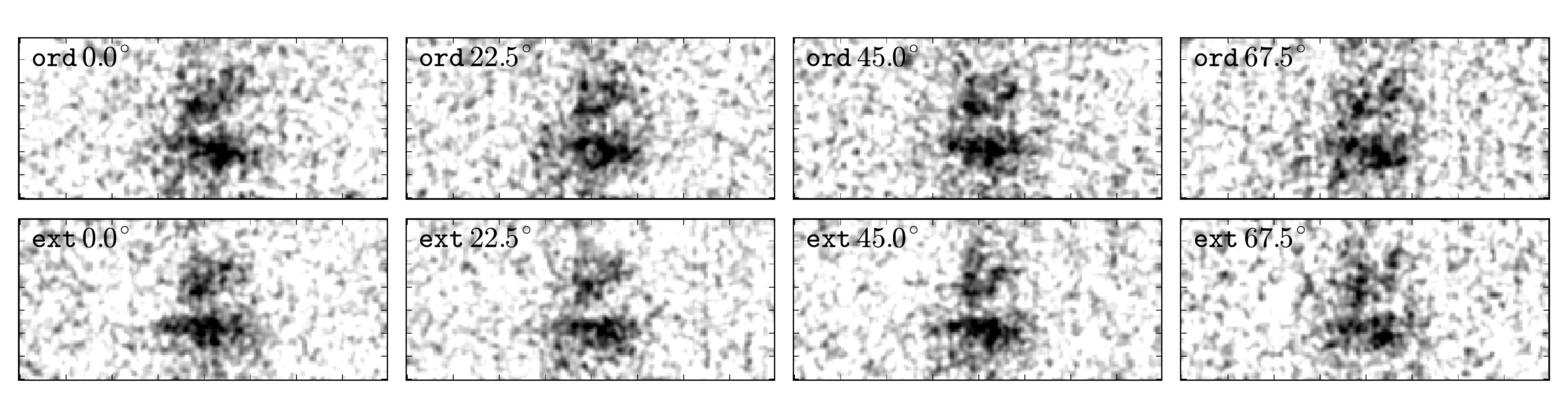}
\caption{Ordinary and extraordinary beams at each HWP angle. Each frame consists of several co-added frames taken over the 5 nights of observations. }
\label{fig: scienceframes}
\end{center}
\end{figure}

In this appendix we demonstrate how our final polarization spectra are created. In \autoref{fig: scienceframes} our eight \ord~and \ext~beams provide the basis of all measurements. These 2D spectra are obtained by stacking their corresponding individual frames as described previously. Combining all eight of these frames yields the total intensity frame shown in \autoref{fig: totint}. It is immediately obvious that there are variations between the \ord~and \ext~beams at each HWP angle which fundamentally leads to a measurement of the polarization.

\begin{figure}[h]
\begin{center}
\includegraphics[width=7.25in]{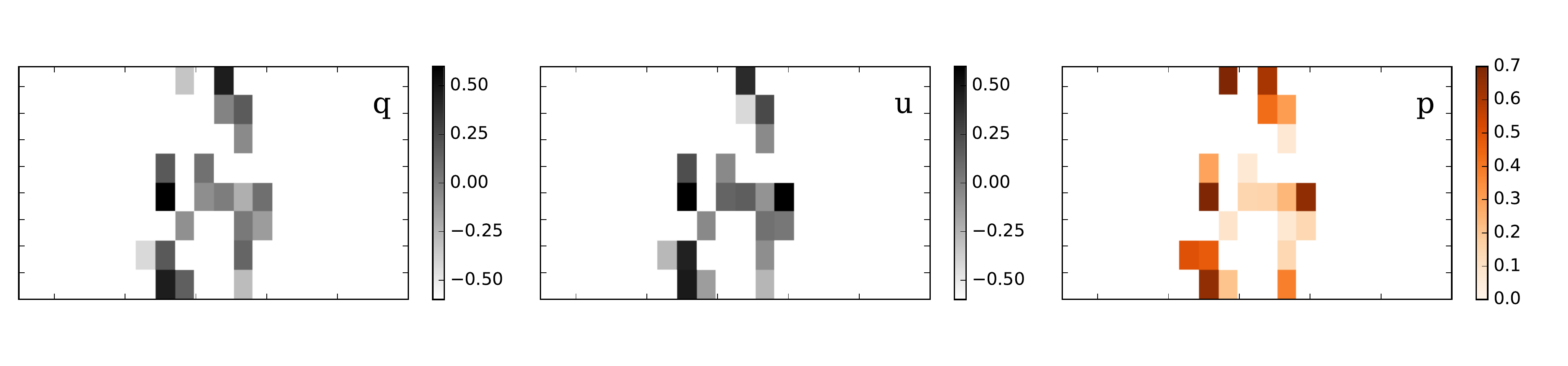}
\caption{2D Stokes parameters and polarization maps. The $q$ and $u$ maps are computed directly from the science frames of \autoref{fig: scienceframes} according to the prescription of \autoref{eqn:stokes} after binning as described in \autoref{sec: pol}. The polarization map in the right panel is then computed via $q$ and $u$ according to \autoref{eqn: pmas}. In all frames, only those bins are shown in which the polarization was deemed significant as discussed in \autoref{sec:analysis}. The variation between the $q$ and $u$ maps is readily seen by eye and is directly related to the amplitude of the polarization measured in that bin.}
\label{fig: QU}
\end{center}
\end{figure}

We then bin these frames according to the various prescriptions described above. From the binned science frames we create Q and U images as shown in \autoref{fig: QU} for the particular case of the 2D binning. In this figure we show only those Q and U bins which correspond to a bin within which we detect a significant polarization signal. One can see by eye the differences in the Q and U frames which is directly responsible for the strength of polarization shown in the third panel. 

\end{document}